\documentclass[preprint]{aastex}

\slugcomment{Submitted to ApJ 2/26/01}

\shorttitle{Heckman et al.}
\shortauthors{Escaping Ionizing Radiation from Starbursts}

\begin{document}


\title{On the Escape of Ionizing Radiation from Starbursts}



\author{T. M. Heckman\altaffilmark{1,2},
K. R. Sembach\altaffilmark{1}, and G. R. Meurer\altaffilmark{1}}
\affil{Department of Physics \& Astronomy, Johns Hopkins University,
Baltimore, MD 21218}

\author{C. Leitherer\altaffilmark{1} and D. Calzetti\altaffilmark{1}}
\affil{Space Telescope Science Institute, Baltimore, MD 21218}

\and

\author{C. L. Martin\altaffilmark{1}}
\affil{Department of Astronomy, Caltech, Pasadena, CA 91125}

\altaffiltext{1}{Guest Investigators on the NASA-CNES-CSA Far
Ultraviolet
Spectroscopic Explorer. FUSE is operated for NASA by the Johns Hopkins
University under NASA contract NAS5-32985.}
\altaffiltext{2}{Adjunct Astronomer, Space Telescope Science Institute}


\begin{abstract}
Far-ultraviolet spectra obtained with $FUSE$ show that the
strong $CII\lambda$1036 interstellar absorption-line is essentially black in
five of the UV-brightest local starburst galaxies.
Since the opacity of the neutral ISM
below the Lyman-edge will be significantly larger than in the $CII$ line,
these data provide strong constraints on the escape of 
ionizing radiation from these starbursts. Interpreted as a
a uniform absorbing slab, the implied optical depth at the Lyman edge is
huge ($\tau_0 \geq 10^2$). Alternatively, the areal covering factor of
opaque material is typically
$\geq$ 94\%.
Thus, the fraction of ionizing stellar photons that escape the
ISM of each galaxy
is small: our conservative estimates typically yield 
$f_{esc} \leq 6\%$. Inclusion of extinction due
to dust will further decrease $f_{esc}$.
An analogous analysis of the rest-UV
spectrum of the star-forming galaxy $MS~1512-CB58$ at $z$ =2.7 leads
to similar constraints on $f_{esc}$.
These new results agree with 
the constraints provided by
direct observations
below the Lyman edge in a few other local starbursts. However, they differ from
the recently reported properties of star-forming galaxies
at $z \geq$ 3.
We assess the idea that 
the strong galactic winds seen in many powerful starbursts clear channels
through their neutral ISM. We show empirically that such
outflows may be a necessary - but not sufficient - part of the
process for creating
a relatively porous ISM. We note that
observations will soon
document the cosmic evolution in the contribution of star-forming galaxies
to the metagalactic ionizing background, with
important implications for the evolution of the IGM.
\end{abstract}


\keywords{galaxies: starburst -- galaxies: intergalactic medium --
galaxies: formation}

\section{Introduction}

The intergalactic medium (IGM) contains the bulk of the baryons in the
universe (e.g., Fukugita, Hogan, \& Peebles 1998).
Determining the source and strength of the metagalactic ionizing radiation field
and documenting its cosmic evolution is crucial to understanding
the fundamental properties of the IGM at both low- and high-redshift.
The two prime candidates for producing the background are QSOs
and star-forming galaxies. While the contribution to the ionizing background
from QSOs can be estimated with reasonable accuracy, considerably
less is known about the contribution from galaxies. QSOs
alone appear inadequate to produce the inferred background
(e.g, Madau, Haardt, \& Rees 1999), especially at $z \geq$ 3
where their co-moving space density declines
steeply with increasing redshift (Fan et al. 2000).
Moreover, there are only rather
indirect constraints on the contribution of galaxies to the ionizing
background in the low-redshift universe (e.g.,
Giallongo, Fontana, \& Madau 1997; Devriendt et al. 1998; Shull et al. 1999).

The cosmic history of the formation of massive (ionizing) stars has
now been determined in broad terms (e.g., Madau, Pozzetti, \& Dickinson
1998; Steidel et al. 1999;
Barger, Cowie, \& Richards
2000). On-average 
several thousand ionizing photons are produced for each nucleon
processed into stars, so the galactic contribution to the
ionizing background is potentially both substantial and calculable. However,
the greatest single uncertainty in this calculation is
the value for $f_{esc}$ - the fraction of the ionizing
photons that escape from star-forming galaxies. 

Uncertainties in the value of $f_{esc}$
also map directly into uncertainties in the use
of standard emission-line diagnostics to infer the basic properties
of star-forming galaxies. Most crucially, the luminosity of
the Hydrogen recombination-lines is a widely utilized
measure of the formation rate of massive stars
in both the local and high-redshift universe (e.g. Gallego et al.
1995; Yan et al. 1999; Pettini et al. 2001). This approach
generally assumes that $f_{esc} << 1$.
However,
the smaller-than-predicted equivalent
widths of the Balmer emission lines in starbursts could imply
that this assumption is suspect
(e.g. Moy et al. 2000; Stasinska et al. 2001).

An $HI$ column of only
1.6 $\times$ 10$^{17}$ cm$^{-2}$ produces $\tau$ = 1 at the Lyman
edge. Mean galactic gas columns are much larger of course, ranging from
$\sim$ 10$^{21}$ cm$^{-2}$ in normal galactic disks to 10$^{24}$ cm$^{-2}$ in
nuclear starbursts (e.g., Kennicutt 1998). Starburst-like mean gas columns
are inevitable in the Lyman Break Galaxies (Heckman 2000), given their
high star-formation rates per unit area (Meurer et al. 1997). The leakage
of ionizing radiation out of galaxies must then be determined by the
topology of the ISM. As such, theory gives us scant guidance, and
so direct measurements of $f_{esc}$ are required.
 
Leitherer et al. (1995) reported the first direct measurements of
$f_{esc}$ using the {\it Hopkins Ultraviolet Telescope} to observe 
far-ultraviolet (far-UV) light
below the rest-frame Lyman edge in a sample of four local starbursts,
and these data were later reanalysed by
Hurwitz, Jelinsky, \& Dixon (1997). The
resulting upper limits on $f_{esc}$ were typically 3\% to 10\%.
In a very surprising development, Steidel, Pettini, \& Adelberger
(2001 - hereafter SPA) have reported
the detection of escaping Lyman continuum in 
the combined spectrum of 29 Lyman Break Galaxies at a mean redshift
$<z>$ = 3.4. They estimate that the ratio of
$f_{esc}$ at 900 \AA\ and 1500 \AA\ ranges from 0.5 to 1 in this composite
spectrum (i.e., there is no definite detection of photoelectric opacity due to
$HI$).
Haehnelt et al.
(2001) reach similar conclusions, but point out that star-forming galaxies at
only slightly smaller redshifts can not be this porous
without violating constraints set by the observed HeII/HI opacity
ratio in the $Ly\alpha$ forest.

It is clearly important to obtain more determinations
of $f_{esc}$, and to use these measurements to better understand the physical
processes that determine $f_{esc}$. This latter goal can be best accomplished
in the local universe, where detailed multiwaveband observations
of star-forming galaxies can be made. Unfortunately, direct observations
of the emerging Lyman continuum require observing
galaxies at redshifts greater than a few percent, so that the foreground
$HI$ opacity of the Milky Way is not significant at the relevant wavelengths
(Leitherer et al. 1995;
Hurwitz, Jelinsky, \& Dixon 1997). Given the modest sensitivity 
of far-UV telescopes, this has limited such investigations to rather
small (and possibly unrepresentative) galaxy samples. 

Against this backdrop, we describe our analysis of new far-UV data obtained
with $FUSE$ for the UV-brightest local starbursts (section 2). 
We point out that the strongest interstellar lines that trace the $HI$ phase
in these starbursts are black (or very nearly black) 
at line-center. We show that
this implies low values for $f_{esc}$ (section 3). This not only more
than doubles the sample of local starbursts with good upper limits
on $f_{esc}$, it illustrates a technique that can potentially be used on a 
much larger sample of galaxies. Indeed, similar
arguments can be applied to high-redshift galaxies (section 4), and
show that $f_{esc}$ is also low in the bright star-forming
galaxy $MS~1512-CB58$ at $z$ =2.7 (Pettini et al. 2000).
We discuss these results and 
their implications in section 5.
 
\section{Observations}

\subsection{The Sample}

We are carrying out a program with the
$Far~Ultraviolet~Spectroscopic~Explorer$
($FUSE$ - Moos et al. 2000) to obtain spectra
of the six starburst galaxies from the Kinney et al. (1993)
`Ultraviolet Atlas of Star-Forming Galaxies' having the largest
UV flux at 1500\,\AA\ through the IUE 
$10\arcsec\times20\arcsec$ aperture (F$_{\lambda}$ $>$
10$^{-13}$ erg cm$^{-2}$ s$^{-1}$ \AA$^{-1}$). These targets span broad ranges
in: 1) metallicity, from 1/8 (NGC~1705) to 2.5 (M~83) times solar,
2) starburst bolometric 
luminosity, from $\sim$3$\times10^8$ (NGC~1705) to $\sim$2$\times10^{10}$
(NGC~3310) L$_{\odot}$,
3) internal dust-reddening, from
A$_V \sim$ 0.0 (NGC~1705) to 0.6 (NGC~3310 and M~83),
and 4) host galaxy properties, from dwarfs (NGC~1705 and NGC~5253), to
irregulars
(NGC~4214 and NGC~4449), to spirals (NGC~3310 and M~83). The sample therefore
spans most of the multidimensional
parameter space of starbursts in the local universe (Heckman et al. 1998).
The selection on the basis of rest-frame $UV$ brightness makes this an
appropriate sample to compare to the UV-selected Lyman Break Galaxies at
high-redshift.

\subsection{Observational Details}
 
The observations undertaken to date are summarized in Table 1. Data
have been obtained for 5 of the 6 targets (only NGC~4449 has not yet
been observed). 
Analyses of the dynamics of the ISM in individual galaxies
based on these data are being reported elsewhere:
Heckman et al. (2001a,b),
Martin et al. (2001).
The large (LWRS, $30\arcsec\times30\arcsec$) aperture on $FUSE$ was used
for the observations of 4 of the 5 targets, while NGC~5253 was observed
with the medium (MDRS, $4\arcsec\times20\arcsec$) aperture.
The corresponding physical sizes of the projected aperture are kpc-scale in
all cases but NGC~5253 (Table 1).

The starburst was centered in the aperture of the
LiF1 (guiding) channel
for each observation by the standard guide-star acquisition procedure.
For four targets, flux was recorded through the LWRS
apertures in both long
wavelength (LiF, $\sim1000-1187$\,\AA) channels and both short wavelength
(SiC, $\sim900-1100$\,\AA) channels. 
We obtained no useful data in the SiC channels during our MDRS
observations of NGC5253 (A0460404) due to a loss of mirror alignment.
We note that while the data do extend slightly below the wavelength
of the starburst's Lyman edge in the SiC1 channel, these data do not usefully
constrain
$f_{esc}$. This is because the opacity of foreground Galactic $HI$ is still
significant
at these wavelengths, largely due to the confluence of the high-order
Lyman series absorption lines (cf. Hurwitz et al. 1997).

Most of the data for this program were obtained at night, which minimizes
\ion{O}{1} and \ion{N}{1} terrestrial airglow contamination in the spectra.  
The percentage of night to total exposure time averaged $\sim65\%$, except for 
NGC~1705 for which the ratio was $\sim92\%$.

The velocity resolution of the data depends on how the far-UV light illuminates
the $FUSE$ apertures. The resolution for a filled $MDRS$ aperture (NGC~5253)
or for a point-like source in the $LWRS$ aperture (NGC~1705) is 
$\sim30$ km~s$^{-1}$. 
We estimate
a resolution of $\leq$50 to $\leq$70 km~s$^{-1}$ in
NGC~3310, M~83, and NGC~4214
based on the widths of the narrowest Galactic (foreground) absorption-lines.

\subsection{Data Processing}

The raw time-tagged photon event lists for each exposure were processed
with the standard $FUSE$ calibration software ($CALFUSE v1.8.5$) available at
the Johns Hopkins University as of January 2001.  The lists were screened
for valid data, and corrections for geometric distortions, spectral motions,
and Doppler shifts were applied (see Sahnow et al. 2000).  The individual
calibrated extracted spectra for each channel were cross-correlated, shifted
to remove residual velocity offsets due to image motion in the apertures,
and combined to produce a composite spectrum.  These composite spectra in
the 1000--1070\,\AA\ region were then compared, and any remaining velocity
offsets were removed by referencing the Galactic absorption lines to the 
appropriate Local Standard of Rest velocities.

Since we are interested in measuring the fluxes in the cores of saturated 
lines, accurate determinations of the residual fluxes are 
essential.  We applied the standard detector background corrections available
within $CALFUSE$ for each observation.  These corrections account only for the 
particle event backgrounds ($\sim0.75$ cnt cm$^{-2}$ s$^{-1}$) and do not 
correct for scattered Ly$\alpha$ emission or other stray light.  These 
additional sources of background light are negligble in the LiF1 data at 
the wavelengths of $\sim1040$\,\AA\ 
(typically $<5\times10^{-15}$ erg cm$^{-2}$ s$^{-1}$ 
\AA$^{-1}$, or less than $\lesssim2\%$ of the continuum flux).  A description
of the FUSE backgrounds can be found in the $FUSE~Observers~Guide$ 
(Blair \& Andersson 2001)\footnotemark.
\footnotetext{This guide (v3.0) can be found on-line at 
{\tt http://fuse.pha.jhu.edu/support/guide/guide.html}.}

In the following discussion
we consider data from the LiF1 channel, which has the highest S/N at the
wavelengths of interest for this study ($\lambda\sim1020-1045$\,\AA).  
The data for the other channels are consistent with our findings presented 
below.

\section{Results from the FUSE Data}

In Figure 1 we show the spectral region around the $Ly\beta$, 
\ion{C}{2} $\lambda$1036.337
and \ion{O}{1} $\lambda$1039.230 absorption-lines in the five galaxies 
in our sample.  
It is immediately apparent that the cores of the \ion{C}{2} $\lambda$1036 and
Lyman lines are quite black. The residual normalized intensities
in the \ion{C}{2} line are typically 
$\leq$ 6\% (Table 2). These \ion{C}{2} residual intensity estimates account 
for overlying Galactic absorption due to the Milky Way \ion{O}{6} 
$\lambda1037.617$ or H$_2$ R(1) $\lambda1037.149$ and P(1) $\lambda1038.157$
lines.  There are no H$_2$ lines within the starbursts themselves that 
contribute to the depth of the \ion{C}{2} line.  We modeled the Galactic 
absorption features using the shorter wavelength member of the \ion{O}{6}
doublet ($\lambda1031.926$) or other H$_2$ R(1) and P(1) lines 
(typically P(1) $\lambda\lambda$1003.294, 1014.326 and 
R(1) $\lambda$1013.435) to get both the velocity extents and expected 
depths of the overlying lines.
We note that the calculation of the
residual intensity is also corrected for the effect of the maximum plausible
stellar contribution to the \ion{C}{2} line. Model starburst
spectra (Gonzalez Delgago, Leitherer, \& Heckman 1997) show that the stellar
photospheric line (which is entirely due to B stars) will depress the local
continuum by a factor of $\sim$ 1.2 to 2 (primarily depending on the age
of the starburst). In our sample, the stellar contribution
to \ion{C}{2} should be largest in the 
B-star-dominated spectrum
of NGC~1705. Even in this case the \ion{C}{2} line is primarily
interstellar, since it (like the other interstellar lines) is blueshifted
by roughly 40 km s$^{-1}$ relative to the galaxy systemic velocity
(Heckman et al. 2001, and see also Heckman \& Leitherer 1997).

\begin{figure}
\plotone{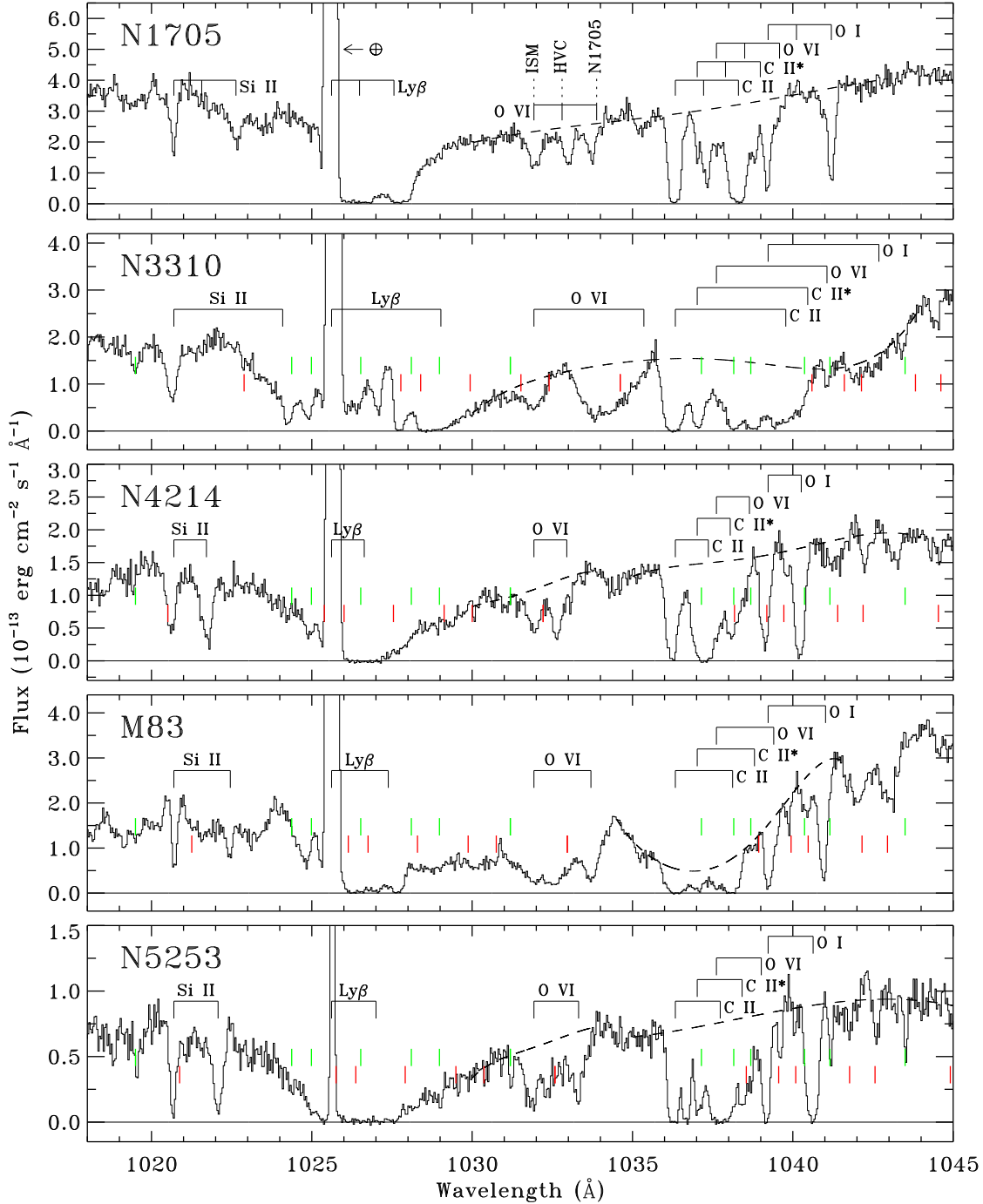}
\vskip -3cm
\caption{$FUSE$ spectra (LiF1A detector segment) of the
spectral region containing the interstellar $Ly\beta$,
\ion{C}{2} $\lambda$1036,
and \ion{O}{1} $\lambda$1039 absorption-lines in five UV-selected local
starbursts.
The lines due to the foreground (Galactic) and starburst ISM are indicated,
as is the fitted continuum (dotted line) to which the residual intensity
at the line core is referenced. Note that the $Ly\beta$ and
\ion{C}{2}\,$\lambda$1036
lines are effectively black in the line core, while the less
saturated \ion{O}{1} $\lambda$1039 line shows a range in residual intensity.
Tick marks indicate the positions (or expected positions) of $J=0-3$
H$_2$ lines in the Milky Way (upper, lighter ticks) and the starbursts
(lower, darker ticks).
All of the spectra shown are a combination of data taken during both
orbital day and night.}
\end{figure}

The \ion{O}{1} $\lambda$1039 line shows behavior
ranging from a normalized residual intensity of about 15\% in NGC~1705
to $\leq$ 6\% in NGC~5253. Overlying Galactic absorption is not an issue
for the \ion{O}{1} at the resolution of the $FUSE$ data.
Both \ion{O}{1} and \ion{C}{2} arise in \ion{H}{1} regions
and contain direct information about the \ion{H}{1} column density
and areal covering factor of the neutral gas.

We begin by writing the ratio of the optical depth at the Lyman edge
to the optical depth at line center. Following Spitzer (1977), this
is

\begin{equation}
\tau_{Ly}/\tau_0 = 4\times10^{-16} (N_H/N_j) (b/f\lambda)
\end{equation}

\noindent
where $(N_H/N_j)$ is the ratio of the \ion{H}{1} and 
ground-state ionic columns,
$b$ is the normal Doppler parameter ($cm~s^{-1}$), $f$
is the oscillator strength, and the wavelength $\lambda$
is measured in $cm$.
Assuming that  \ion{O}{1} and \ion{C}{2} are the dominant ionic species of 
their respective element in the \ion{H}{1} phase, this can be written as

\begin{equation}
\tau_{Ly}/\tau_0 = R Z_{gas}^{-1} (b/100~km/s) 
\end{equation}

\noindent
where $Z_{gas}$ is the gas-phase $C$ or $O$ abundance in solar units and
$R$ = 8.6 for \ion{C}{2} $\lambda$1036 and $R$ = 50 for \ion{O}{1}
$\lambda$1039
(Morton 1991). The important point is that {\it the optical depth at
the Lyman edge is much larger than the optical depth at line-center}
for either line for typical values of $Z_{gas}$ and $b$. Corrections for 
grain-depletion will only increase the implied ratio of $\tau_{Ly}/\tau_0$.
Ionization corrections are unimportant for \ion{O}{1} since it is closely
coupled to \ion{H}{1} through charge exchange reactions, and will only 
increase the implied ratio of $\tau_{Ly}/\tau_0$ for \ion{C}{2}.

As an example, consider the case of NGC~1705 (Heckman et al. 2001).
Based on the nebular emission-lines, the gas-phase Oxygen abundance
is 1/8 solar, and this is consistent 
with the abundances derived in the \ion{H}{1} gas for $Si$, $Ar$, and $Fe$
from the $FUSE$ spectra. The
measured $b$ value is 40$\pm$10 km s$^{-1}$. The residual intensity
at the core of the $OI\lambda$1039 line implies $\tau_0$ = 1.8
(for unit covering factor - see below). Equation 2 above then
implies that $\tau_{Ly}$ = 280. This is roughly consistent with the $HI$
column we derived from fitting the $Lyman$ series lines:
$N_{HI}$ = 1.5$\times$ $10^{20}$ cm$^{-2}$. Thus, in the
context of a homogeneous
absorbing slab, the escaping fraction of ionizing photons 
will be almost identically zero in NGC~1705.
In principle, even more severe constraints on
$\tau_{Ly}$ can be placed using weaker absorption-lines
(transitions having smaller values
of $(N_j/N_H)f\lambda$). We do not take this approach, since even
the strong \ion{O}{1} $\lambda$1039 line already implies $f_{esc}\sim$0.

Less stringent (more realistic) constraints on $f_{esc}$
can be placed
by adopting a ``picket fence'' model
in which the areal covering factor of optically-thick $HI$
is not unity (e.g., ionizing radiation can escape through ``holes''
in the $HI$). In this case, the upper limit on the residual intensity
at the center of the $CII\lambda$1036 line implies that the
covering factor must be $\geq$ 94\%, so that the upper bound
on the escape fraction for ionizing radiation is $f_{esc} \leq$ 6\%.

We have undertaken this analysis for the other four galaxies in
our sample, and report the results in Table 2. In all cases, the
results are similar to those in NGC~1705, with implied values
for $f_{esc}$ that are almost identically zero (for the 
homogeneous slab model) and $\leq$6\% for the picket-fence model.
The values listed in Table~2 should be considered stringent upper limits
for the following reasons: 1) We list only the 
limits derived from the picket fence model. 
2) We measured $f_{esc}$ at velocities where the 
residual flux was greatest after accounting for possible contamination 
by other absorption features.  3) Scattered light may contribute at low 
levels ($\lesssim2\%$) and only serves to strengthen the limit on $f_{esc}$.
4) The FUSE line spread function (LSF) likely consists of two components: a 
narrow component accounting for $\sim80\%$ of the LSF area and a broad 
component accounting for $\sim20\%$ and having $\sim2-3$ times the width of the
narrow component.  While the exact details of the LSF
are presently unknown, such a model more accurately describes the 
shapes of the cores of the Galactic \ion{H}{1} lines toward nearby stars 
than a single component LSF.  As a result, it is possible that some of the 
observed flux in the line cores is simply redistributed light from nearby 
continuum regions within $\sim1$\,\AA\ of the line.  5) The continuum
estimates for the interstellar lines shown in Figure 1 are conservative
estimates and are unlikely to be lower than those shown.
6) The 
upper limits on $f_{esc}$ can be pushed to
even smaller values by including the opacity due
to dust (that is, our technique measures only the contribution
of the photoelectric opacity of \ion{H}{1}). Application of the
empirical starburst dust attenuation law (Calzetti 1997;
Meurer, Heckman, \& Calzetti 1999) to the $IUE$ spectra
of these galaxies implies an extinction due to dust at 1500\AA\ ranging
from negligible (NGC~1705) to $\sim$ 3 magnitudes in
M~83 and NGC~3310. The mean dust opacity in the Lyman continuum 
(1 to 4 Ryd) should be larger still (Mathis 1990).

We conclude that our UV-selected starbursts are quite opaque to
their own ionizing radiation. This confirms the results of
Leitherer et al. (1995) and Hurwitz, Jelinsky, \& Dixon (1997), but
generalizes and strengthens them.

\section{Galaxies at High-Redshift}

The technique described above can be applied to galaxies at high-redshift.
As an application, we consider the 
star-forming galaxy $MS~1512-CB58$ at $z$ =2.7 (Yee et al. 1996).
Thanks to strong gravitational lensing, its flux has been
amplified by a factor of $\sim$ 30 (Seitz et al. 1998). Thus, the 
rest-frame UV spectrum 
discussed by Pettini et al. (2000) is by far the best of its kind.
We display the spectral region from $Ly\alpha$ to $CII\lambda$1335
in Figure 2. The quoted spectral resolution in these data is
$\sim$200 km s$^{-1}$ FWHM, while the interstellar lines are much broader
(500 to 700 km s$^{-1}$) and very deep.
It is instructive to note that the value
for $f\lambda$ (equation 1) for the $CII\lambda$1335 ($OI\lambda$1302)
line is 0.1 (0.8) dex
larger than for the $CII\lambda$1036 ($OI\lambda$1039) line in our $FUSE$ 
spectra. Thus, the spectrum of $MS~1512-CB58$ can
be straightforwardly compared to the local starbursts.

\begin{figure}
\plotone{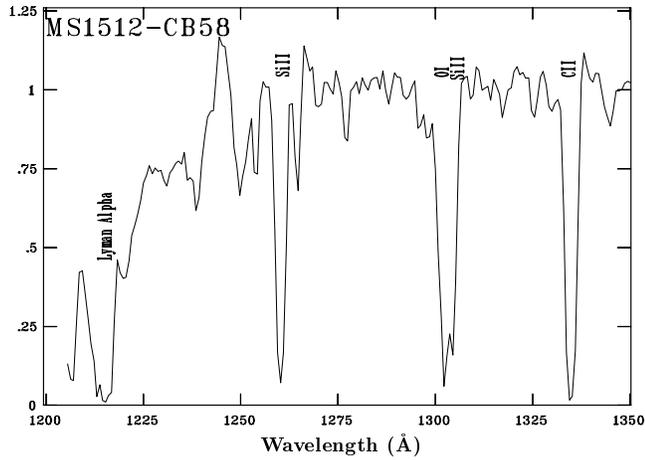}
\vskip -8cm
\caption{A portion of the $Keck$ LRIS spectrum of the
gravitationally-lensed
star-forming galaxy $MS~1512-CB58$ at $z =$2.7 published by
Pettini et al. (2000). The interstellar $Ly\alpha$, \ion{Si}{2} $\lambda$1260,
\ion{O}{1} $\lambda$1302, \ion{Si}{2} $\lambda$1304, and
\ion{C}{2} $\lambda$1335 absortion-lines
are indicated. The stellar continuum in the spectrum has been normalized
to unit intensity by fitting it with a low-order polynomial. Note that the
interstellar lines are nearly black, and much broader than the
instrumental resolution of 0.8 \AA\ (rest-frame).}
\end{figure}

The measured residual intensites in the line cores in the 
$MS~1512-CB58$ spectrum are: $Ly\alpha$ (1\%),
$SiII\lambda$1260 (7\%), $OI\lambda$1302 (6\%), and
$CII\lambda$1335 (2\%).
To apply equations 1) and 2) above, we adopt
a gas-phase metallicity of 1/4 solar
(Pettini et al. 2000;
Leitherer et al. 2001; Tremonti et al. 2001)
and take $b$ = 300 km s$^{-1}$ based
on the line-widths reported by Pettini et al. (2000).
The implied optical depth at the Lyman edge assuming a uniform
slab is $\tau_{Ly} \sim$ 300. Pettini et al. (2000) derive an $HI$
column of 7.5$\times$10$^{20}$ cm$^{-2}$ from the damped $Ly\alpha$
line, yielding an even larger value for $\tau_{Ly}$ for a uniform slab.
Alternatively, the residual intensity
at the core of the $CII\lambda$1335 line implies an areal
covering factor of optically-thick gas of 98\%. The implied
value for $f_{esc}$ ranges from $\sim$0 to 2\%. These values
are consistent with the results for the local starbursts in section 3 above.

The situation is quite different in the faint Lyman Break Galaxies
at $z \geq$ 3. As SPA point out,
the interstellar
lines in their composite spectrum are far from black (the
apparent residual intensities are roughly 70\% at line center). The
composite spectrum of Lyman Break Galaxies at $<z>$ = 3.0 in
the Hubble Deep Field is similar (Lowenthal et al. 1997). It is possible
in these cases that spectra with higher spectral resolution and signal-to-noise
would show that the lines are in fact nearly black over narrow ranges
in velocity. However, the existing data are consistent with the
conclusion of SPA that the escape
fraction of ionizing radiation is large in these galaxies.

\section{Discussion}

The results we have reported for five of the six UV-brightest local starbursts
(NGC~1705, NGC3310, NGC~4214, NGC~5253, and M~83) implies
that UV-bright starburst galaxies today are highly opaque to their own ionizing
radiation. This agrees with results for a small sample of more distant
starbursts described by Leitherer
et al. (1995) and Hurwitz et al. (1997). These results in turn are consistent
with the estimate by Shull et al. (1999) that the strength of the 
local metagalactic 
ionizing background implies that the fraction of ionizing photons 
that escape from
star-forming galaxies can not exceed a global average value of about 5\%.

These results are in striking contrast to the detection of significant
escaping Lyman continuum radiation from Lyman Break Galaxies
at $z \sim$ 3 to 4 by SPA. Our analysis
shows that the star-forming galaxy $MS1512-CB58$ at $z=$2.7
is similar to the local starbursts: it too must be highly opaque below
the Lyman break.
Taking the current results at face value, we are left with a puzzle.
Why is the topology/structure of the neutral interstellar medium
apparently so different between the galaxies studied by 
SPA and those in the present paper?

In some ways this is a surprising result, because
local starbursts and
Lyman Break Galaxies are remarkably similar in most respects
(e.g., Heckman 2000).
Meurer et al. (1997)
showed that both have the same
bolometric surface-brightnesses, and therefore the same star formation rates
per unit area.
This implies that
the basic physical properties of the ISM should be similar. 
The surface-mass-densities in the stars ($\Sigma_*$) and the
interstellar gas ($\Sigma_g$)
that fuels the star-formation must be similar. 
The radiant energy density in the ISM must be similar
($\sim$10$^3$ times higher than in the Milky Way), as must
the rate of mechanical energy deposition
(supernova heating) per unit area or volume.
Simple considerations of hydrostatic equilibrium imply correspondingly
high total pressures in the ISM in both:
$P \sim G \Sigma_g
\Sigma_{tot}$, or $P/k \sim$ 
few $\times$ 10$^7$ K cm$^{-3}$. This is several thousand times the value
in the local ISM in the Milky Way. Finally, the
characteristic dynamical time in the ISM
will be short in both cases: $t_{dyn}$
$\sim$ $(G\rho)^{-1/2}$ $\sim$ $(G\Sigma_{tot}H)^{-1/2}$
$\sim$ few Myr (where $H \sim$ 10$^2$ pc is the thickness of the disk).
Perhaps the most striking difference is the size-scale: starbursts
in the local universe are generally circum-nuclear events (sizes of $10^2$ to
$10^3$ pc) imbedded in a much larger galaxy, while the intense star-formation
in the Lyman Break
Galaxies appears to be a galaxy-wide event (sizes of $10^3$ to $10^4$ pc).
This might lead to fundamental differences in the global properties of the ISM.

Such generic arguments aside, is there any empirical evidence that
the structure or dynamics in the ISM in our specific sample
of local starbursts differs from the Lyman Break Galaxies? We
consider two possibilities.

First, our $FUSE$ sample galaxies
have considerably lower luminosities (star-formation rates)
than the typical Lyman Break Galaxies: $logL_{bol} \sim$ 8.5 to 10.3
$L_{\odot}$ {\it vs.} 10.7 to 11.7 $L_{\odot}$ (Adelberger \& Steidel 2000).
It seems physically plausible that the ISM might become more
porous in starbursts with higher star-formation (and energy deposition)
rates (cf. Lehnert \& Heckman 1996).
We have tested this idea by considering data on $f_{esc}$
for more powerful local starbursts. Two of the four starbursts observed
below the Lyman break by Leitherer et al. (1995) are powerful
systems with $L_{bol} \sim 10^{11} L_{\odot}$. Both are opaque in the
Lyman continuum:
$f_{esc} \leq$ 2\% and $\leq$ 4\% for Mrk~496 and IRAS~08339+6517 respectively
(Hurwitz et al. 1997).
We have also searched the HST
archives for UV spectra of powerful local starbursts taken with
a spectral resolution comparable to our $FUSE$ data and with a projected
aperture size ($\sim$kpc-scale) comparable to our $FUSE$ data. There are
three such starbursts in the sample investigated by Kunth et al. (1998), who
used the $GHRS$ echelle mode to observe the $Ly\alpha$, $OI\lambda$1302,
and $SiII\lambda$1304 lines with a resolution of $\sim$ 20 km s$^{-1}$.
These starbursts have luminosities ($L_{\odot}$) of
$logL_{bol}$ = 10.4 (ESO~400-G043), 11.0 (IRAS~08339+6517), and 11.2
(ESO~350-IG038). These interstellar lines are black in two cases
(ESO~400-G043 and ESO~350-IG038), and have
a residual intensity
of 60 to 70\% in the third case (IRAS~08339+6517). While a large value
for $f_{esc}$ is therefore {\it possible} in this galaxy, direct
observations below the Lyman break actually show that $f_{esc} \leq$ 4\%
in this case (Hurwitz et al. 1997). 

A second possibility is that large values for $f_{esc}$ are created
by supernova-driven galactic outflows 
(superwinds) that clear out channels through which ionizing radiation
can escape (SPA). The tracers of superwinds in Lyman Break Galaxies are their
blueshifted interstellar absorption-lines (Franx et al. 1997; 
Pettini et al. 1998; 2000; 2001).
Indeed, the composite
spectrum of the 29 Lyman Break Galaxies analyzed by SPA
shows the clear signature of these outflows: the interstellar
absorption-lines are blueshifted by 400 to 500 km s$^{-1}$ with
respect to the $Ly\alpha$ emission-line. This outflow signature is also seen in
the composite spectrum formed from the sum of the spectra
of 12 Lyman Break Galaxies in the HDF at $z \sim$ 3 (Lowenthal et al. 1997; 
Franx et al. 1997).
As noted above, the relative
shallowness of the interstellar lines in both these composite spectra
is consistent with (but does not demand)
a large value for $f_{esc}$.

Similar outflows are seen in the UV spectra of local starbursts
(Kunth et al. 1998; Gonzalez Delgado et al. 1998), and
Heckman et al. (2000) have shown that high-velocity outflows of
neutral gas are common in powerful ($L_{bol} \geq$ 10$^{11}$ $L_{\odot}$)
local starbursts. In contrast, our $FUSE$ data
reveal high-velocity ($\Delta$$v \geq$ 10$^2$ km s$^{-1}$) outflows in the
neutral phase of the ISM in only one of our five
galaxies: NGC~3310, in which the mean outflow  
velocity is about -200 km s$^{-1}$ (see Figure 1).
In the other galaxies the neutral-phase absorption-lines 
arise in relatively quiescent gas near the galaxy
systemic velocity (Figure 1; Martin et al. 2001; Heckman et al. 2001).
Based on the ISM dynamics, it is perhaps not surprising that
NGC~1705, NGC~4214,
NGC~5253, and M~83 are opaque to their Lyman continuum radiation.

On the other hand, Gonzalez Delgado et al. (1998) have shown that
outflows at several hundred km s$^{-1}$ are present in the neutral
gas in at least three of 
the four starbursts
observed below the Lyman break by Leitherer et al. (1995). In all
cases the galaxy is opaque below the Lyman break, with $f_{esc}$ typically
$\leq$ 3 to 10\%. Of the three powerful starbursts observed by
Kunth et al. (1998), two (ESO~400-G043 and ESO~350-IG038) show
black interstellar lines and outflows at $(v - v_{sys})$ =
-225 and -58 km s$^{-1}$
respectively. In the third (IRAS~08339+6517), 
the lines have a residual intensity
of 60 to 70\%, and are strongly blueshifted by $\sim$-500
km s$^{-1}$ (Gonzalez Delgado et al. 1998).

We conclude that the currently avalilable data do not demonstate
that galactic winds inevitably produce large values of $f_{esc}$
in local starbursts.
Such outflows appear to be a necessary - but not sufficient -
part of the process that creates an ISM that is porous to ionizing radiation.

In the near future we will use
$FUSE$ to observe a sample of more powerful starbursts
($L_{bol}$ = 4$\times$10$^{10}$ to 4$\times$10$^{11}$ $L_{\odot}$).
This will allow us to probe the Lyman continuum opacity in more starbursts
with luminosities (and star formation rates) that are similar to those
of typical Lyman Break Galaxies.
At the same
time, the extension of the investigation at high-redshift
to a wider range of Lyman Break
Galaxies can help clarify the situation. In particular,
it will be important to test the high values for
$f_{esc}$ by obtaining data on both the Balmer recombination-lines
and the Lyman continuum in the same set of galaxies (to provide a
consistency check).
Finally, the
$Galex$ mission (Martin et al. 1999) will directly measure the escaping
flux below the Lyman break for the population of star-forming galaxies in the
redshift range $z \sim$ 0.4 to 2 (the epoch that apparently dominates the 
cosmic history
of star-formation).
Thus, over the next few years, we should be able to make the first 
good direct measurement
of the contribution of star-forming galaxies to the metagalactic
ionizing background as a function of cosmic epoch. This will have
important implications for the history of the IGM.

\acknowledgments

We thank the members of the $FUSE$ team for providing this superb
facility to the astronomical community. We thank Max Pettini
for enlightening discussions, and for kindly providing us
a copy of the composite Lyman Break Galaxy spectrum published
by SPA.
This work was supported in part
by NASA grants NAG5-6400 and NAG5-9012.  KRS acknowledges support from
NASA Long Term Space Astrophysics grant NAG5-3485.

\clearpage

\begin{deluxetable}{lcccccccc}
\tablecolumns{0}
\tablewidth{0pt}
\tablecaption{FUSE Starburst Observations}
\tablehead{Galaxy & $l$ & $b$ & $F_{1050\,\AA}$ & Dataset & Obs. Date
 & T$_{exp}$\tablenotemark{a} & Aper.\tablenotemark{b} & $L$\tablenotemark{c}\\
&($\degr$)&($\degr$)&(erg\,cm$^{-2}$\,s$^{-1}$\,\AA$^{-1}$)&&(y/m/d)&(ksec)&&
(kpc)}
\startdata
NGC~1705 & 261.08 & $-$38.74 & $\sim4.2\times10^{-13}$ & A0460102 & 2000/02/04 &
 \phn7.7 & LWRS & 0.9$\times$0.9 \\
         &        &          &                         & A0460103 & 2000/02/05 &
 13.6 & LWRS\\
\\
NGC~3310 & 156.01 & +54.06 & $\sim2.6\times10^{-13}$   & A0460201 & 2000/05/05 &
 27.1 & LWRS & 2.56$\times$2.56\\
\\
NGC~4214 & 160.25 & +78.08 & $\sim1.6\times10^{-13}$   & A0460303 & 2000/05/12 &
 20.7 & LWRS & 0.50$\times$0.50\\
\\
M~83     & 314.58 & +31.97 & $\sim3.0\times10^{-13}$   & A0460505 & 2000/07/06 &
 26.5 & LWRS & 0.55$\times$0.55\\
\\
NGC~5253 & 314.86 & +30.11 & $\sim8.9\times10^{-14}$   & A0460404 & 2000/07/07 &
 27.4 & MDRS & 0.40$\times$0.08\tablenotemark{d}\\
\enddata
\tablenotetext{a}{On-target exposure time in the LiF1 channel.}
\tablenotetext{b}{Aperture sizes.  MDRS:~4$\times$20 $arcsec$;
LWRS:~30$\times$30 $arcsec$. }
\tablenotetext{c}{Physical size of projected FUSE aperture.}
\tablenotetext{d}{Aperture position angle = 195.87$\degr$.  No useful
SiC data obtained for this MDRS observation.}
\end{deluxetable}

\clearpage

\begin{deluxetable}{lccl}
\tablecolumns{4}
\tablewidth{0pt}
\tablecaption{Residual Intensities in the Core of the C II 1036.337\,\AA\ Line}
\tablehead{Galaxy & $v_{sys}$ & $f_{esc}$(C II) & \multicolumn{1}{c}{Notes on $f
_{esc}$ and nearby absorption}\\
&        (km~s$^{-1}$) & (\%)\\}
\startdata
NGC~1705 & 569 & $\le 5.8$ & Weak Galactic high velocity \ion{O}{6} $\lambda1037
.617$ in red wing of profile.\\
& & & No H$_2$ absorption at nearby velocities. \\
\\
NGC~3310 & 997 & $\le 6-30$ & Value of $f_{esc}$ increases along redward wing of
 absorption.\\
\\
NGC~4214 & 298 & $\le 4.0$ & Weak Galactic H$_2$ R(1) $\lambda1037.149$ on blue
side of profile.\\
\\
M~83 & 517 & $\le 6.0$ &
Weak Galactic H$_2$ P(1) $\lambda1038.157$ on red side of profile.\\
& & & Galactic \ion{O}{6} $\lambda1031.926$ on blue side of profile. \\
\\
NGC~5253 & 405 & $\le 6.0$ & Weak Galactic H$_2$ P(1) $\lambda1038.157$ on red s
ide of profile.\\
& & & Galactic \ion{O}{6} $\lambda1031.926$ on blue side of profile. \\
\enddata
\end{deluxetable}


\end{document}